%% file: ssvep-gan.tex
\documentclass[conference]{IEEEtran}
\IEEEoverridecommandlockouts

\usepackage{cite}
\usepackage[pdftex]{graphicx}
\graphicspath{{../pdf/}{../jpeg/}}
\DeclareGraphicsExtensions{.pdf,.jpeg,.png}
\usepackage[cmex10]{amsmath}
\usepackage{amssymb}
\usepackage{algorithmic}
\usepackage[autostyle]{csquotes}
\usepackage{caption}
\usepackage{subcaption}
\usepackage{url}
\usepackage{array}
\usepackage{mdwmath}
\usepackage{mdwtab}
\usepackage{eqparbox}
\usepackage{booktabs} 

\usepackage[breaklinks=true,bookmarks=false,hyperfootnotes=false]{hyperref}

\renewenvironment{thebibliography}[1]{
  \begin{oldthebibliography}{#1}
    \setlength{\itemsep}{0.0em}
    \setlength{\parskip}{0.0em}
}
{
  \end{oldthebibliography}
}
\begin{document}
%
\title{Simulating Brain Signals: Creating Synthetic EEG Data via Neural-Based Generative Models for Improved SSVEP Classification}

\author{\IEEEauthorblockN{Nik Khadijah Nik Aznan\IEEEauthorrefmark{1}\IEEEauthorrefmark{2},
Amir Atapour-Abarghouei\IEEEauthorrefmark{1},  Stephen Bonner\IEEEauthorrefmark{1}, Jason D. Connolly\IEEEauthorrefmark{3}\\
Noura Al Moubayed\IEEEauthorrefmark{1} and Toby P. Breckon\IEEEauthorrefmark{1}\IEEEauthorrefmark{2}}

\IEEEauthorblockA{ Department of $\{$\IEEEauthorrefmark{1}Computer Science, \IEEEauthorrefmark{2}Engineering, 
\IEEEauthorrefmark{3} Psychology$\}$ \\ Durham University, Durham, UK }
}


\maketitle

\begin{abstract}
Despite significant recent progress in the area of Brain-Computer Interface (BCI), there are numerous shortcomings associated with collecting Electroencephalography (EEG) signals in real-world environments. These include, but are not limited to, subject and session data variance, long and arduous calibration processes and predictive generalisation issues across different subjects or sessions. This implies that many downstream applications, including Steady State Visual Evoked Potential (SSVEP) based classification systems, can suffer from a shortage of reliable data. Generating meaningful and realistic synthetic data can therefore be of significant value in circumventing this problem. We explore the use of modern neural-based generative models trained on a limited quantity of EEG data collected from different subjects to generate supplementary synthetic EEG signal vectors, subsequently utilised to train an SSVEP classifier. Extensive experimental analysis demonstrates the efficacy of our generated data, leading to improvements across a variety of evaluations, with the crucial task of cross-subject generalisation improving by over 35\% with the use of such synthetic data.
\end{abstract}


%
\IEEEpeerreviewmaketitle

\input{Sections/intro.tex}
\input{Sections/lit_review.tex}
\input{Sections/method.tex}
\input{Sections/experiments.tex}
\input{Sections/result.tex}
\input{Sections/conclusion.tex}



\bibliographystyle{IEEEtran}
\bibliography{paper_ref}

\end{document}

%% file: Sections/intro.tex
\section{Introduction}
\label{sec:intro}

Electroencephalography (EEG) is the most prominent signal acquisition approach employed for Brain-Computer Interface (BCI) as it has the non-invasive ability to capture electrical activity of the human cerebral cortex \cite{rao2013brain}. BCI is a system that translates such acquired signals to provide a communication and control medium between the human brain and external devices. BCI has received significant attention within the research community for decades \cite{lotte2018review}. However, not many BCI applications are tractable for daily use in real-world scenarios, especially for important medical applications, such as assisting patients with locked-in syndrome. This is due to the numerous shortcomings and limitations within the current state of the art that lead to the low reliability and usability of BCI \cite{lotte2011generating}.

\begin{figure}[!t]
    \centering
    \includegraphics[width=0.80\linewidth]{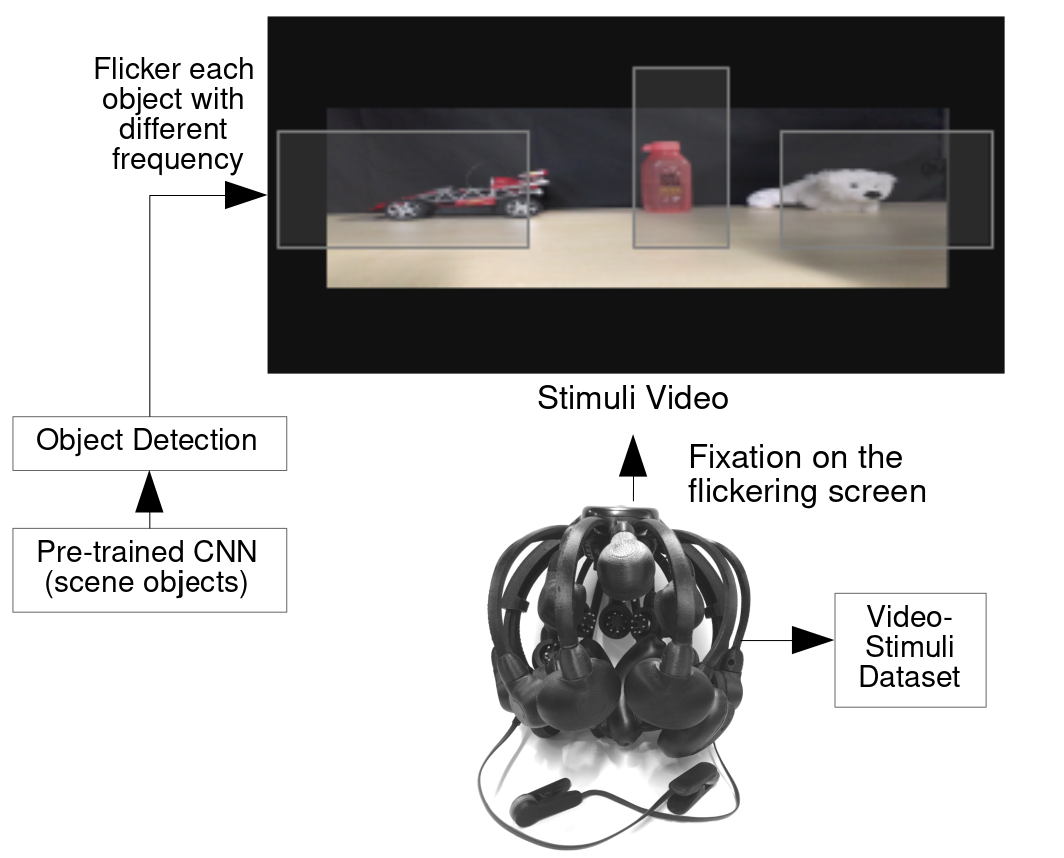} 
    \caption {Details of the collection process for an SSVEP EEG dataset (\emph{Video-Stimuli}) using a video based stimuli.}
    \vskip -20pt
    \label{fig:Gen_ds}
\end{figure}

Deep neural networks have recently been used to improve the classification of various aspects of EEG data \cite{kwak2017convolutional, aznan2018classification, thomas2017deep}. While it is essential to have access to large quantities of data for training such methods, collecting high quality EEG data has proven difficult \cite{zhang2018improving, luo2018eeg}. This can be for a variety of reasons including the requirement for careful per-subject and per-session calibration. This makes EEG BCI experiments time-consuming, expensive and difficult to operate within the usually short amount of time experimental subjects can perform EEG experiments \cite{corley2018deep}. In addition to these issues, there are commonly-known limitations with EEG data in general, which can severely hinder the applicability of a system dependent on such data \cite{zhang2018improving}. For instance, EEG data is known to be highly subject and session variant, which leads to a long calibration process for every individual experiment \cite{lotte2011generating, lotte2018review}. This further impacts any machine learning based models which are trained upon this data, as they often demonstrate poor generalisation performance across different subjects or experiential data collecting sessions \cite{lotte2011generating, thomas2017deep}.

In machine learning, generative models have long been used to generate entirely new and realistic data points which match the distribution of a given target dataset \cite{salakhutdinov2015learning}. Recent work on neural-based models such as Generative Adversarial Networks (GAN) and Variational Auto-Encoders (VAE) have demonstrated that these are highly capable at capturing key elements from a diverse range of datasets to generate realistic samples \cite{goodfellow2014generative}. Increasingly, there is evidence that using synthetic data, taken from a generative model, can be used as a form of data augmentation to help improve the performance of any down-stream data classification task \cite{AdarDKAGG18}.

In this paper, we detail the generation of new synthetic EEG data using a selection of customised neural-based generative models and explore applications of such data including using it to boost classification accuracy on real datasets. Uniquely, we specifically focus on dry-EEG data containing Steady State Visual Evoked Potential (SSVEP) signals. Dry-EEG requires no conductive gel which improves its usability within a BCI context, eliminating major limitations of the wet EEG systems \cite{edlinger2012can, Lopez-Gordo2014, minguillon2017trends}. However, dry-EEG results in high impedance values that cause more noise and artefacts in the data, leading to more challenging signal decoding and classification \cite{edlinger2012can}. 

SSVEP is a type of evoked potential stimuli generated by having repeated flashes at certain frequencies presented to subjects (the flashing can occur in a video as seen in Figure \ref{fig:Gen_ds}) \cite{rao2013brain}. The frequency of the flashing is present in the EEG signals recorded from the subjects and can be extracted via a variety of competing signal processing techniques \cite{liu2014recent}. SSVEP has many important applications in BCI, for example it can be utilised to allow people with severe physical disabilities to control or communicate with external devices just by having them fixate on a flickering stimuli \cite{dehzangi2018portable}, controlling an exoskeleton \cite{kwak2017convolutional} or navigating a humanoid robot \cite{aznan2018using}. To the best of our knowledge, this is the first work in the literature to explore the use of neural-based generative models to create dry-EEG data containing SSVEP information. In summary, we make the following major contributions in this work: 

\begin{itemize}
    \item The generation of synthetic dry-EEG data containing SSVEP signals using a variety of unsupervised models.
    \item A demonstration that using generated data can improve the classification of real-world EEG data, taken from multiple subjects and recorded under various conditions and sessions.
    \item An exploration of both classifier pre-training and dataset augmentation as use cases for the generated data. 
    \item Further demonstration that using synthetic EEG data can increase the convergence rate of classification models, thus resulting in the observation that smaller quantities of real-world training data is required. 
\end{itemize}

We perform extensive experiential evaluations to validate our claims and to aid reproducibility, we release our Python-based (PyTorch) \href{https://github.com/nikk-nikaznan/SSVEP-Neural-Generative-Models}{implementation} of all the generative models, along with sample input data\footnote{\href{https://github.com/nikk-nikaznan/SSVEP-Neural-Generative-Models}{https://github.com/nikk-nikaznan/SSVEP-Neural-Generative-Models}}.

%% file: Sections/lit_review.tex
\section{Related Work}
\label{sec:related-work}

We consider the background information relevant to this work within two distinct areas:- recent advances made in generative models (Section \ref{ssec:generative-models}) and existing work on using such models to generate meaningful and coherent synthetic EEG data (Section \ref{ssec:literature-review}).

\subsection{Neural-Based Generative Models}
\label{ssec:generative-models}

Generative models have been proven very powerful within the context of unsupervised learning where the model learns a hidden structure of the data from its distribution to generate new data samples within the same distribution \cite{jaakkola1999exploiting}. This generated dataset often contains enough variation to support the down-stream training of a secondary model \cite{luo2018eeg}.

Generative Adversarial Networks (GAN) \cite{goodfellow2014generative} are capable of producing semantically sound artificial samples by inducing a competition between a generator ($G$), which attempts to capture the distribution, and a discriminator ($D$), which assesses the generator output and penalizes unrealistic samples. Both networks are trained simultaneously to achieve an equilibrium.

Training a GAN is known to be challenging with pervasive instability issues \cite{arjovsky2017towards}. One such issue stems from the discriminator rapidly reaching optimality and effortlessly distinguishing between the fake samples output by the generator and samples from the real distribution. This will lead to a lack of meaningful gradients for training, effectively ceasing any progress towards the equilibrium.

The Wasserstein GAN (WGAN) is consequently proposed in \cite{arjovsky2017wasserstein} to rectify some of the issues associated with training a GAN. The Wasserstein-1 metric is used to measure the distance between the real and model distributions. Also known as the Earth Mover's distance, ($EM(p,q)$), this metric is the minimum cost of moving distribution elements (earth mass) to transform a distribution $q$ to distribution $p$ (cost = mass $\times$ transport distance).

The Wasserstein GAN \cite{arjovsky2017wasserstein} has an aptly named \emph{critic} ($C$) instead of a discriminator since this network is no longer a classifier. Using the EM distance, the critic will not only determine whether a sample is fake or real as a discrete binary decision, but how real or how fake the generated sample is as a continuous regressive output. Under the right training circumstances, the critic will eventually converge to a linear function with ever-present meaningful gradients and cannot saturate. The loss function in the Wasserstein GAN is created via the Kantorovich-Rubinstein duality \cite{arjovsky2017wasserstein}:\vspace{-0.05cm}
\begin{equation}  
    \min_{G} \max_{C \in \mathcal{F}}\ \mathop{\mathbb{E}}_{x\sim \mathbb{P}_{r}} [C(x)] - \mathop{\mathbb{E}}_{\tilde{x} \sim \mathbb{P}_{g}} [C(\tilde{x})],
    \label{eq:W-GAN}\vspace{-0.25cm}
\end{equation}\\
where $\mathcal{F}$ is the set of 1-Lipschitz functions, $\mathbb{P}_{r}$ the real distribution, $\mathbb{P}_{g}$ the model distribution defined by $\tilde{x} = G(z), z \sim p(z)$, and $z$ the random noise. If $C$ is optimal, minimizing the value function with respect to $G$ minimizes $EM(\mathbb{P}_{r},\mathbb{P}_{g})$.

The Wasserstein GAN does not suffer from vanishing gradients or mode collapse. However, to guarantee continuity, a Lipschitz constraint must be enforced, which is achieved in \cite{arjovsky2017wasserstein} by \emph{clamping} the weights. This results in the creation of a new hyper-parameter, which needs to be carefully tuned to the distribution.

A gradient norm penalty with respect to the critic input is consequently proposed in \cite{gulrajani2017improved} to replace clamping. Since a differentiable function is 1-Lipschitz if and only if its gradient norm is no more than 1 everywhere, \cite{gulrajani2017improved} limits the critic gradient norm by penalizing the function on the gradient norm for samples $\hat{x} \sim \mathbb{P}_{\hat{x}}$, where $\hat{x} = \epsilon x + (1 - \epsilon)\tilde{x},\ 0 < \epsilon < 1$. This penalty term which is added to the function in Eqn. \ref{eq:W-GAN} is therefore as follows \cite{gulrajani2017improved}:
\begin{equation}
    \centering
        \mathop{\mathbb{E}}_{\hat{x} \sim \mathbb{P}_{\hat{x}}} [(||\triangledown_{\hat{x}} C(\hat{x})||_{2} - 1)^2],
    \label{eq:penalty-improved-W-GAN}
\end{equation}\\
where $\mathbb{P}_{\hat{x}}$ is implicitly defined to sample uniformly along straight lines between pairs of points sampled from the real data distribution, $\mathbb{P}_{r}$, and $\mathbb{P}_{g}$ is the model distribution defined by $\tilde{x} = G(z), z \sim p(z)$ \cite{gulrajani2017improved}.

Auto-encoders have long been used as a method of creating a low-dimensional representation $z$ of data using an encoder model, which can be used to reconstruct the original data with minimal errors via a decoder model \cite{hinton2006reducing}. However, traditionally they cannot be explicitly used to generate new data samples based on the learned data distribution. Variational Auto-encoders (VAE) utilise ideas from Bayesian inference to produce a more expressive data representation, whilst also having the ability to generate new data samples \cite{kingma2013auto, rezende2014stochastic}. Unlike non-probabilistic auto-encoders \cite{hinton2006reducing}, a VAE does not learn a fixed value for each element in $z$ but instead each element is sampled from a probability distribution before being passed to the decoder model. This has been shown to produce a more semantically meaningful representation, where individual dimensions in the hidden space can correspond to tangible elements in the dataset, such as facial expression in a dataset of human faces \cite{larsen2015autoencoding}.

As the decoder model of the VAE is trained to take a sample from a Gaussian distribution and produce a realistic output, it can be used to produced new data by simply sampling points in the distribution and reconstructing them.

\subsection{Literature Review}
\label{ssec:literature-review}

Within the existing body of work in the literature, there are instances of generative models used to create synthetic EEG data. For instance, \cite{hartmann2018eeg} proposes using the improved WGAN with GP for a single channel of the EEG data for motor imagery task. The generator consists of one linear layer, six convolutional layers where each layer consists of an up-sampling operation, two convolutions and one fully-connected layer. The authors evaluate the performance of four models, three different up-sampling methods (nearest-neighbour, linear interpolation and cubic interpolation with convolutional down-sampling) and down-sampling via average pooling with the original WGAN-GP. All the models are then assessed using four evaluation metrics; inception score, Frechet inception distance, Euclidean distance and sliced Wasserstein distance. All four models are demonstrated to outperform WGAN-GP. 

In \cite{corley2018deep}, the authors propose deep EEG super-resolution using a GAN. The model is applied to a small number of EEG channel data to interpolate other channel signals using motor imagery dataset from \cite{millan2004need}. The super-resolution data (SR) is generated via WGAN with convolutional filters using the low-resolution data (LR) from the dataset by down-sampling the EEG channels by scale factors of two and four for two different experiments. The channels removed from the down-sampling processes are used as high-resolution data (HR) to compete against SR in the discriminator. They evaluate the performance of the SR by performing classification and comparing the accuracy with the classification performed using HR. The authors conclude that SR is capable of producing high spatial resolution EEG signals from low resolution signals. 

Instead of generating EEG signals, \cite{palazzo2017generative} generates previously seen images while having brain signals recorded by EEG. Six subjects are shown 2000 images with 40 classes per subject. The EEG signals are pre-processed by hardware notch filter between 49 and 51 Hz and bandpass filter between 14 and 70 Hz. Using LSTM RNN, the temporal feature representations are encoded, which are subsequently used as condition vectors employed along with random noise vectors to generate new images. They obtain a test accuracy of 83.9\% when evaluating the LSTM model for feature representation. 

In \cite{zhang2018improving}, synthetic EEG signals are generated by inverting the artificial time-frequency representation (TFR) obtained from conditional Deep Convolutional GAN (cDCGAN). The Wavelet Transform is used to obtain the TFR of the signals to be used by the cDCGAN to generate the artificial TFR of the EEG signals. The BCI competition II dataset III \cite{schlogl2003outcome} is used as the EEG data. They evaluate the efficacy of the synthetic data by comparing the classification accuracy of the model trained on real data, synthetic data and a mixture of real and synthetic data with different ratios. Using additional synthetic data, the accuracy improved up to 3$\%$.

In \cite{luo2018eeg} and \cite{qiu2018data}, a GAN is used to generate synthetic emotion recognition EEG data. The authors in \cite{luo2018eeg} generate EEG in the form of differential entropy from noise distribution using a conditional WGAN with two emotion recognition datasets. They evaluate the performance of the generated data by combining the synthetic and real-world data to train a classification model compared against a model solely trained on real-world data. The addition of the synthetic data leads to an improvement in the accuracy of up to 20$\%$ for different datasets.

The approach in \cite{qiu2018data} combines classification and generative networks in a model using an Auxiliary Classifier GAN (AC-GAN). Part of the approach is encoding the data from SEED and DEAP datasets into images using a Markov Transition Field (MTF) before passing them into the AC-GAN to generate new data samples. Every sample comes with a corresponding class label. A Tiled CNN is employed to classify the MTF images to either fake or real and with the class label. They improved the classification by less than 1$\%$ in the SEED dataset as compared with previous work.

In this paper, we generate synthetic EEG signals from SSVEP tasks using dry-EEG. While earlier work \cite{zhang2018improving, qiu2018data} has been able to generate features within a secondary domain, from which EEG signals are subsequently reconstructed, we are the first to introduce the concept of generating meaningful EEG data directly in signal space via end-to-end training instead of first transforming the signals into different domains for SSVEP classification. This improves overall efficiency, removes any need for additional signal processing, and prevents potential instability errors introduced during transformations. 

%% file: Sections/method.tex
\section{Proposed Approach}
\label{sec:approach}

As one of the objectives of this work is to create synthetic EEG data, which can be used to improve the training of a downstream classifier \cite{aznan2018classification}, we explore the use of a variety of neural-based generative models (as described in Section \ref{sec:related-work}). In this section, a brief description of the details of the models used in this work is provided.

\subsection{Deep Convolutional Generative Adversarial Network}
\label{sec:approach:GAN}

As part of this work, we use the generative model proposed in \cite{radford2015unsupervised}. Random noise vectors $z$ are sampled from a Gaussian distribution and used as the generator input. At every iteration, the generator outputs fake data samples ($\tilde{x} = G(z)$), which are then passed to the discriminator along with randomly selected real data samples $x$, classifying them as either fake or real, the gradients from which are employed to train the generator, leading to higher quality outputs at every step. An overview of our generative adversarial model is seen in Figure \ref{fig:GAN}.

\subsubsection{Loss Function}
\label{sec:approach:GAN:loss}

\begin{figure}[!h]
  \centering
  \includegraphics[width=0.9\linewidth]{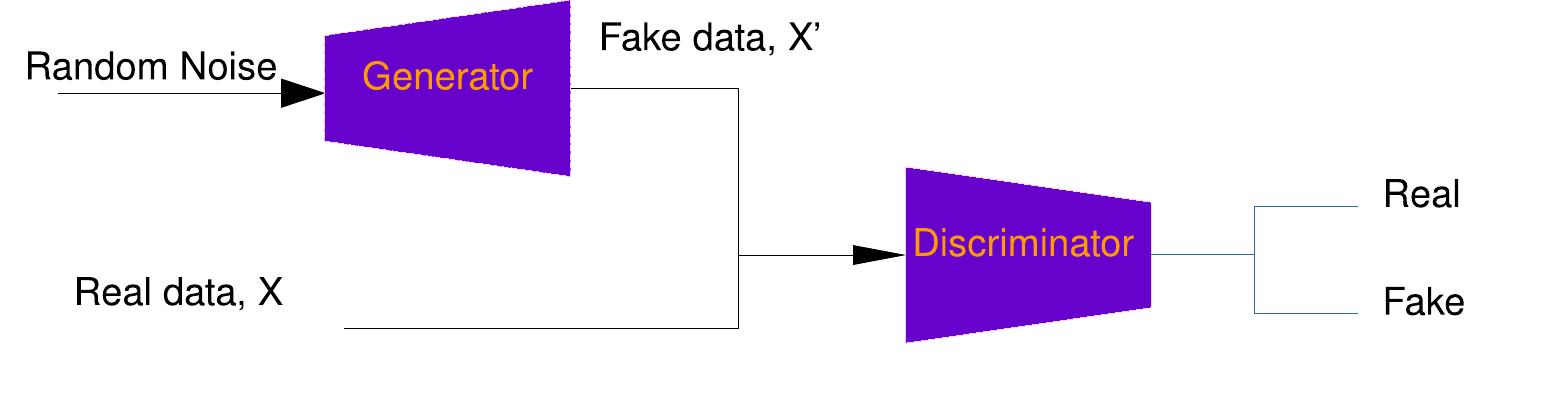} 
  \caption{An overview of the generative adversarial network.}
  \vskip -10pt
  \label{fig:GAN}
\end{figure}

The loss function is based on the competition between the generator and the discriminator following the minimax objective \cite{goodfellow2014generative}:
\begin{equation}  
    L =\min_{G} \max_{D}\ \mathop{\mathbb{E}}_{x\sim \mathbb{P}_{r}} [log(D(x))] + \mathop{\mathbb{E}}_{\tilde{x} \sim \mathbb{P}_{g}} [log(1 - D(\tilde{x}))],
    \label{eq:vanilla-GAN}
\end{equation}
where $\mathbb{P}_{r}$ is the data distribution, $\mathbb{P}_{g}$ the model distribution defined by $\tilde{x} = G(z), z \sim p(z)$, and $z$ the random noise vector used as the input to the generator. An overview of the training pipeline is seen in Figure \ref{fig:GAN}.

\subsubsection{Implementation Details}
\label{sec:approach:GAN:Implementation}

The network architecture is based on that of \cite{radford2015unsupervised}, with the exception of the use of one dimensional convolutions since the networks process EEG signal vectors rather than images. Our generator consists of one dense layer and three 1D transpose convolutional layers.

This vector is then used as the input to our light-weight discriminator, consisting of two layers; the first containing a convolution-BatchNorm-LeakyReLU($slope=0.2$) module followed by max pooling, and the second using a fully-connected layer followed by a leaky ReLU ($slope=0.2$). 

Implementation and training is carried out in \emph{PyTorch} \cite{pytorch}, with Adam \cite{kingma2014adam} used as the optimization approach (momentum $\beta_{1} = 0.5$, $\beta_{2} = 0.999$, initial learning rate $\alpha = 0.0001$).

\subsection{Improved Wasserstein Generative Adversarial Network} 
\label{sec:approach:W-GAN}

Similarly to the training procedure in Section \ref{sec:approach:GAN}, the Wasserstein GAN is made up of two completing networks, a generator and a critic. The generator receives a random noise vector $z$ as its input, and the critic determines how real or fake the data samples created by the generator are by calculating the distance between the real data distribution and the model distribution (Earth Mover's distance).

Since it is significantly important to keep the critic optimal at all times, we train the critic 25 times per each generator training iteration for the first 100 generator iterations and 5 times per each generator iteration for the rest of the training process. 

\subsubsection{Loss Function}
\label{sec:approach:W-GAN:loss}

Here, we take advantage of the improved Wasserstein GAN \cite{gulrajani2017improved} with the following loss function:
\begin{equation}
  \centering
  \begin{split}  
      L = \min_{G} \max_{C} & \mathop{\mathbb{E}}_{\tilde{x} \sim \mathbb{P}_{g}} [C(\tilde{x})] - \mathop{\mathbb{E}}_{x\sim \mathbb{P}_{r}} [C(x)] + \\
      & \lambda \mathop{\mathbb{E}}_{\hat{x} \sim \mathbb{P}_{\hat{x}}} [(||\triangledown_{\hat{x}} C(\hat{x})||_{2} - 1)^2],
      \label{eq:improved-W-GAN}
  \end{split}
\end{equation}
where $\mathbb{P}_{g}$ is the model distribution defined by $\tilde{x} = G(z), z \sim p(z)$, $z$ is random noise, $\mathbb{P}_{r}$ is the true data distribution, and $\mathbb{P}_{\hat{x}}$ is implicitly defined to sample uniformly along straight lines between pairs of points sampled from $\mathbb{P}_{r}$ and $\mathbb{P}_{g}$.

\subsubsection{Implementation Details}
\label{sec:approach:W-GAN:Implementation}

For the sake of consistency, the architecture of the networks used here are similar to that of the networks in Section \ref{sec:approach:GAN}. Similarly, all implementation and training is carried out in \emph{PyTorch} \cite{pytorch}, with Adam \cite{kingma2014adam} used as the optimization approach (momentum $\beta_{1} = 0.1$, $\beta_{2} = 0.999$, initial learning rate $\alpha = 0.0001$).

\subsection{Variational Autoencoder}
\label{sec:approach:VAE}

In addition to the GAN based approaches, we also explore the creation of a Convolutional Variational Auto-encoder (detailed in Section \ref{ssec:generative-models}) to generate synthetic EEG data. Our VAE model uses 1D convolutions to encode features from a given EEG data sample, which are used to parametrise a Gaussian distribution from which a latent and compressed representation of the input data is sampled. This latent representation is passed to the decoder section of the model which comprises transposed convolutions used to transform the latent representation back into the original EEG data. 

\begin{figure}[!h]
  \centering
  \includegraphics[width=1.0\linewidth]{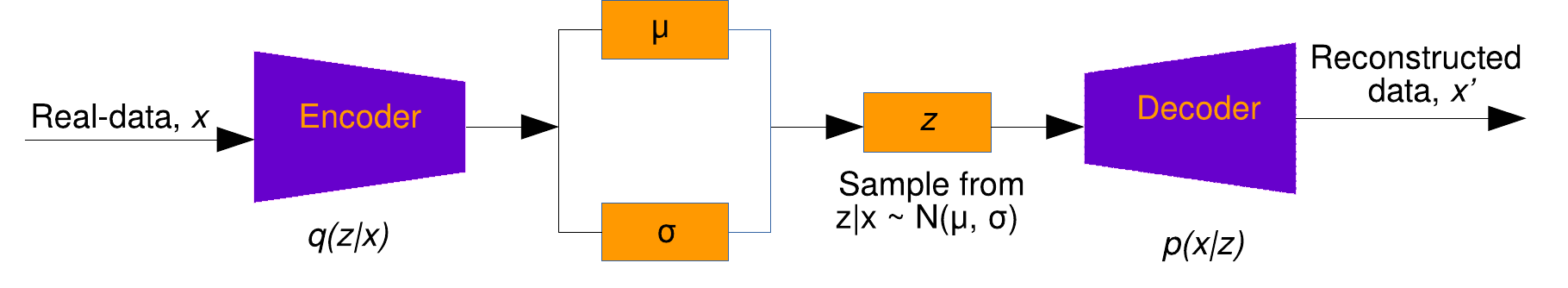} 
  \caption{An overview of the Variational Auto-encoder.}
  \vskip -10pt
  \label{fig:VAE}
\end{figure}

Once we have trained the VAE model using the objective function detailed in the next section, we are able to use it to generate entirely new data samples, which could have plausibly come from the training data, but which is not conditional on any input to the model.

\subsubsection{Loss Function}
\label{sec:approach:VAE:loss}

The encoder section of our VAE model is trained to learn two output vectors, $\mu$ and $\sigma$, which represent the mean and variance of the latent space from which $z$ will be sampled, $z = \mathcal{N}(\mu, \sigma)$. Using the sampled representation $z$, the decoder section of the VAE is trained to reconstruct the input data $x$. Consequently, our VAE is trained to infer the intractable distribution $p(z|x)$, that being the likelihood of $z$ given the data $\hat x$, using a stand-in tractable one $q(z|x)$ in the following manner:
\begin{equation}
      L = \mathbb{E}_{q(z|x)} \Big[ log (p(x|z)) \Big] - KL(q(z | x) || p(z)),
      \label{eq:VAE-learn}
  \end{equation}
where $KL$ is the Kullback-Leibler distance between $p$ and $q$, $q(z|x)$ is the output of the convolutional based encoder portion of our VAE and $p(x|z)$ is the output from the decoder section. We make use of a Gaussian prior as the distribution for $p(z)$. Figure \ref{fig:VAE} shows the the general layout of our VAE model.

\subsubsection{Implementation Details}
\label{sec:approach:VAE:Implementation}

We utilise a convolutional encoder for our VAE model and a transpose convolution based decoder. The encoder consists of a 1D convolution, with BatchNorm and leaky ReLU ($slope=0.2$) as the activation function, followed by max pooling. This common learned feature representation is passed into two separate linear layers which learn the $\mu$ and $\sigma$ used to parametrise the Gaussian and generate $z$. The decoder architecture comprises three stacked 1D transpose convolutional layers all using leaky ReLU ($slope=0.2$) to perform the reconstruction from $z$ to create $\hat x$.

As with the GAN models, all implementation and training is carried out in \emph{PyTorch} \cite{pytorch}, with Adam \cite{kingma2014adam} used as the optimization algorithm (momentum $\beta_{1} = 0.1$, $\beta_{2} = 0.999$, initial learning rate $\alpha = 0.0001$).

%% file: Sections/experiments.tex
\section{Experimental Setup}
\label{sec:experimental-setup} 

This section will detail the setup of our experimental evaluation, including introducing the empirical datasets used, detailing how we generate new data samples from our generative models and our procedure for evaluating the quality of the generated data.

\subsection{Empirical SSVEP Dry-EEG Datasets}
\label{sec:empirical-datasets}

\begin{figure}[!h]
  \centering
  \includegraphics[width=0.8\linewidth]{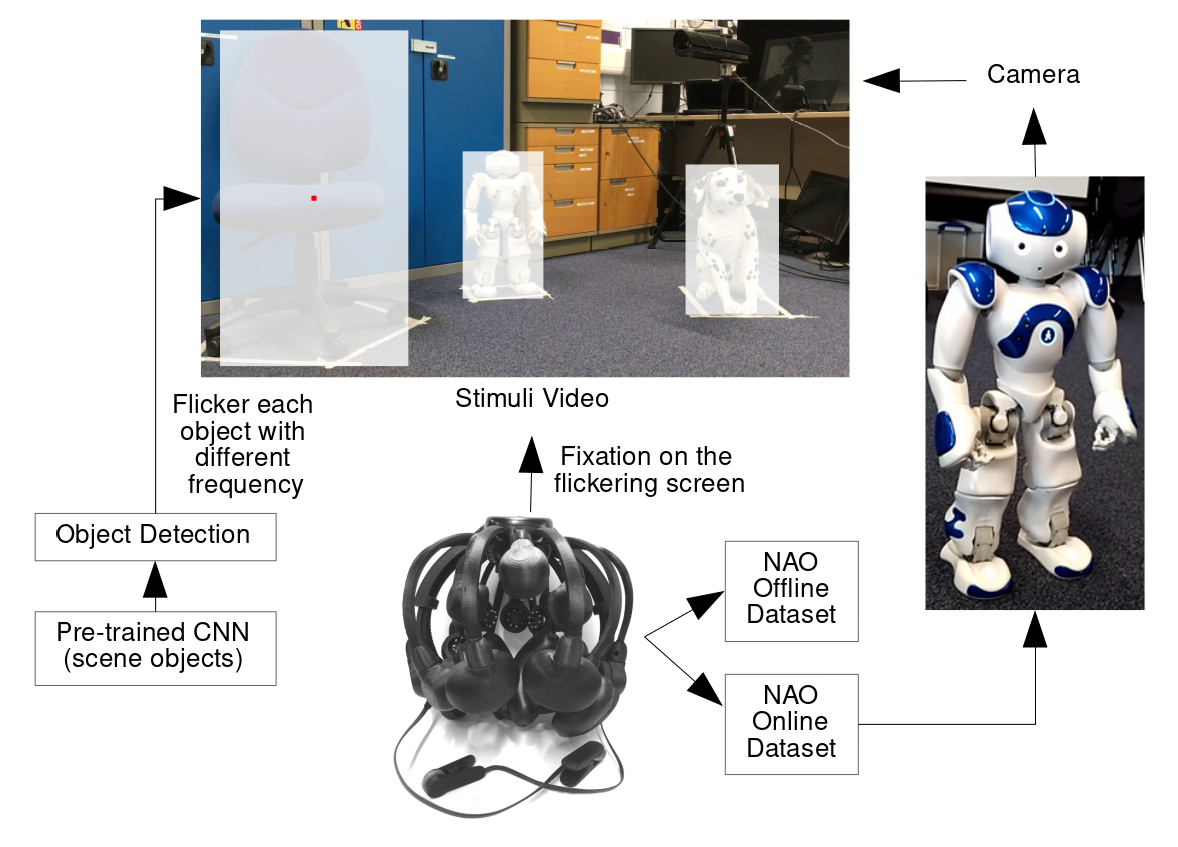} 
  \caption{Flowchart detailing the recording procedure for the \emph{Online} and \emph{Offline} SSVEP EEG Datasets. The highlighted region of the figure shows the humanoid robot used to create the \emph{Online dataset}.}
  \vskip -5pt
  \label{fig:dataset}
\end{figure}  

We make use of two empirical SSVEP dry-EEG datasets which we collected and fully detailed in our previous work \cite{aznan2018using}. The collection procedure for this dataset is detailed in Figure \ref{fig:dataset}. This empirical data is used as a way of validating that the generated data is realistic enough to be used to improve the performance of a classification model. In this dataset, objects are detected in a video sequence using a pre-trained object detection model \cite{liu2016ssd}. The detected objects are then flickered by rendering black/white polygon boxes on top of the objects with display frequency modulations of 10, 12 and 15 Hz to create the frequencies, common to both datasets, which we attempt to detect via the SSVEP paradigm. Our two empirical datasets are detailed below:

\begin{itemize}

  \item \emph{Video-Stimuli Dataset:} This dataset is collected from an offline experiment (shown in Figure \ref{fig:Gen_ds}) and is used as a basis for our generative models to learn the distribution of the SSVEP data. This dataset comprises 50 unique samples recorded for each of the three classes, taken from one subject (S01) performing the SSVEP task by looking at objects detected in a pre-recorded video sequence. 

  \item \emph{NAO Dataset:} This dataset comprises data from an experiment containing both offline and online elements (highlighted in Figure \ref{fig:dataset}). The offline portion of the data contains 50 unique samples for each of the three classes taken from three subjects (S01, S02, S03) performing the SSVEP task by looking at objects detected in a pre-recorded video sequence from a humanoid robot. The online portion of the data contains 30 samples per class taken from the same three subjects when navigating the robot in real time \cite{aznan2018using}.
\end{itemize}

\begin{figure}[!h]
  \centering
  \includegraphics[width=0.8\linewidth]{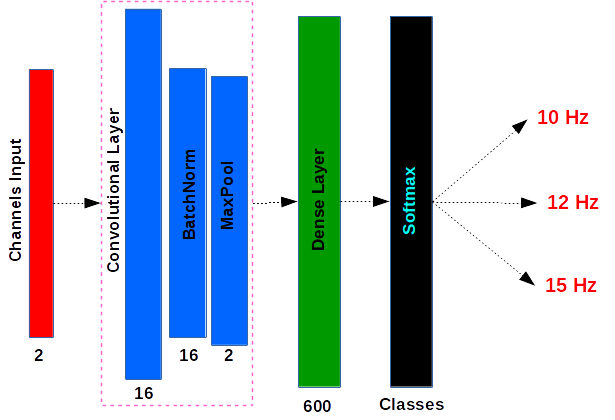} 
  \caption{The CNN architecture used to classify the EEG signals.}
  \vskip -10pt
  \label{fig:CNN}
\end{figure}

\begin{figure*}[!h]
  \centering
  \begin{subfigure}[b]{0.31\textwidth}
    \includegraphics[width=1.0\linewidth]{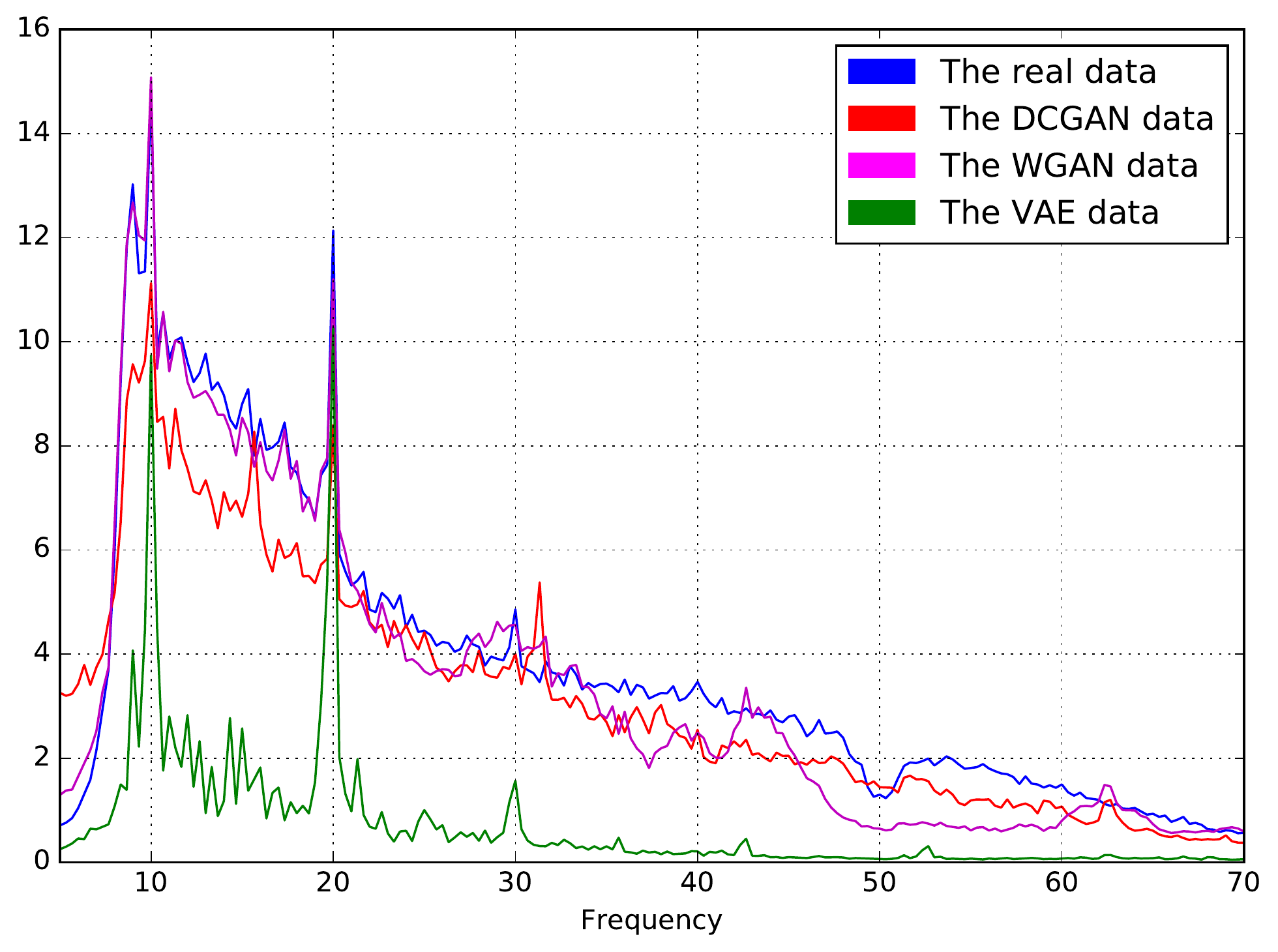}
    \vspace{-0.5cm} 
      \caption{SSVEP Signal at 10Hz.}
      \label{fig:class0}
  \end{subfigure}
  \hfill
  \begin{subfigure}[b]{0.31\textwidth}
      \includegraphics[width=1.0\linewidth]{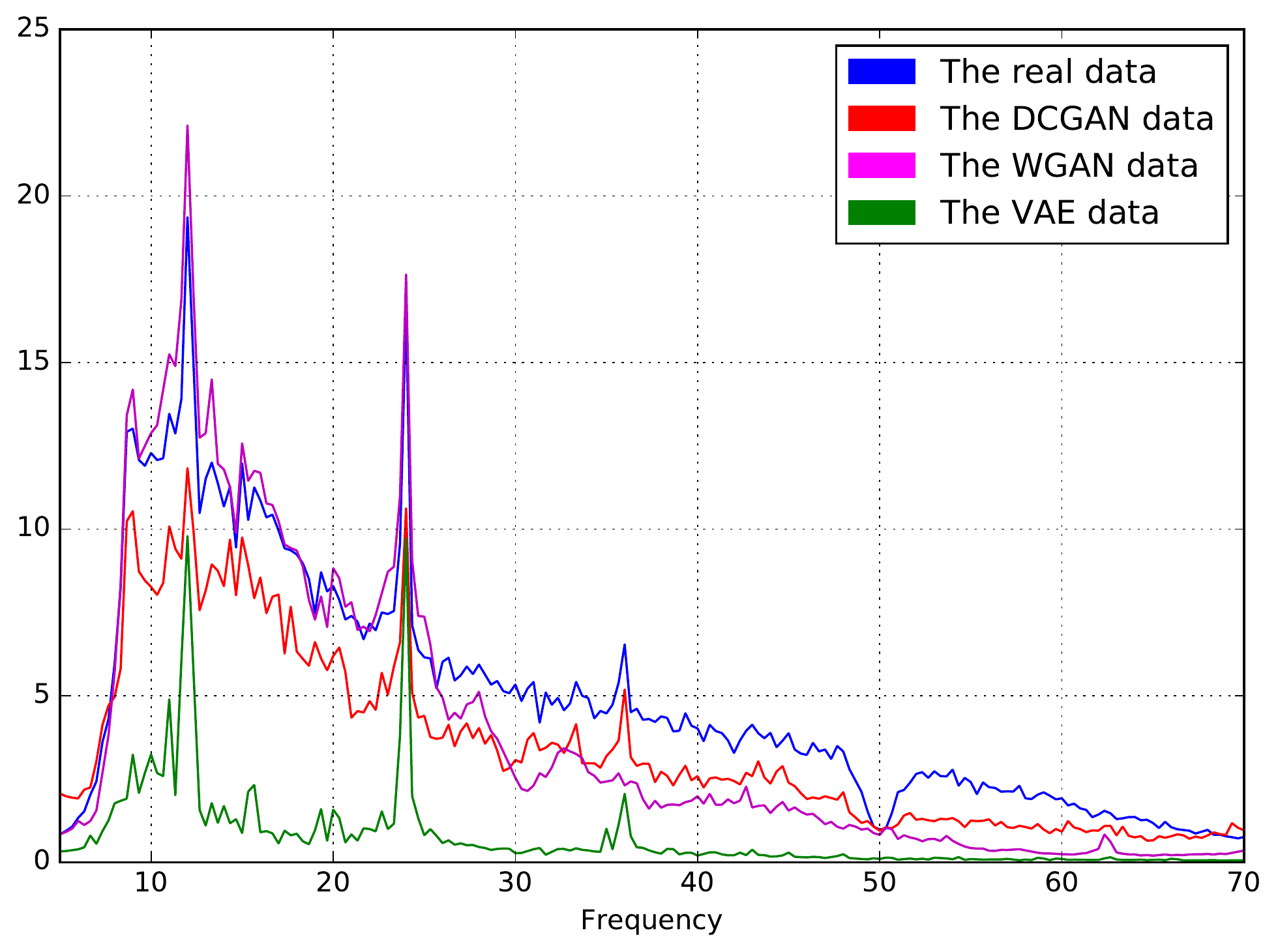}
      \vspace{-0.5cm} 
      \caption{SSVEP Signal at 12Hz.}
    \label{fig:class1}
  \end{subfigure}
  \hfill
  \begin{subfigure}[b]{0.31\textwidth}
      \includegraphics[width=1.0\linewidth]{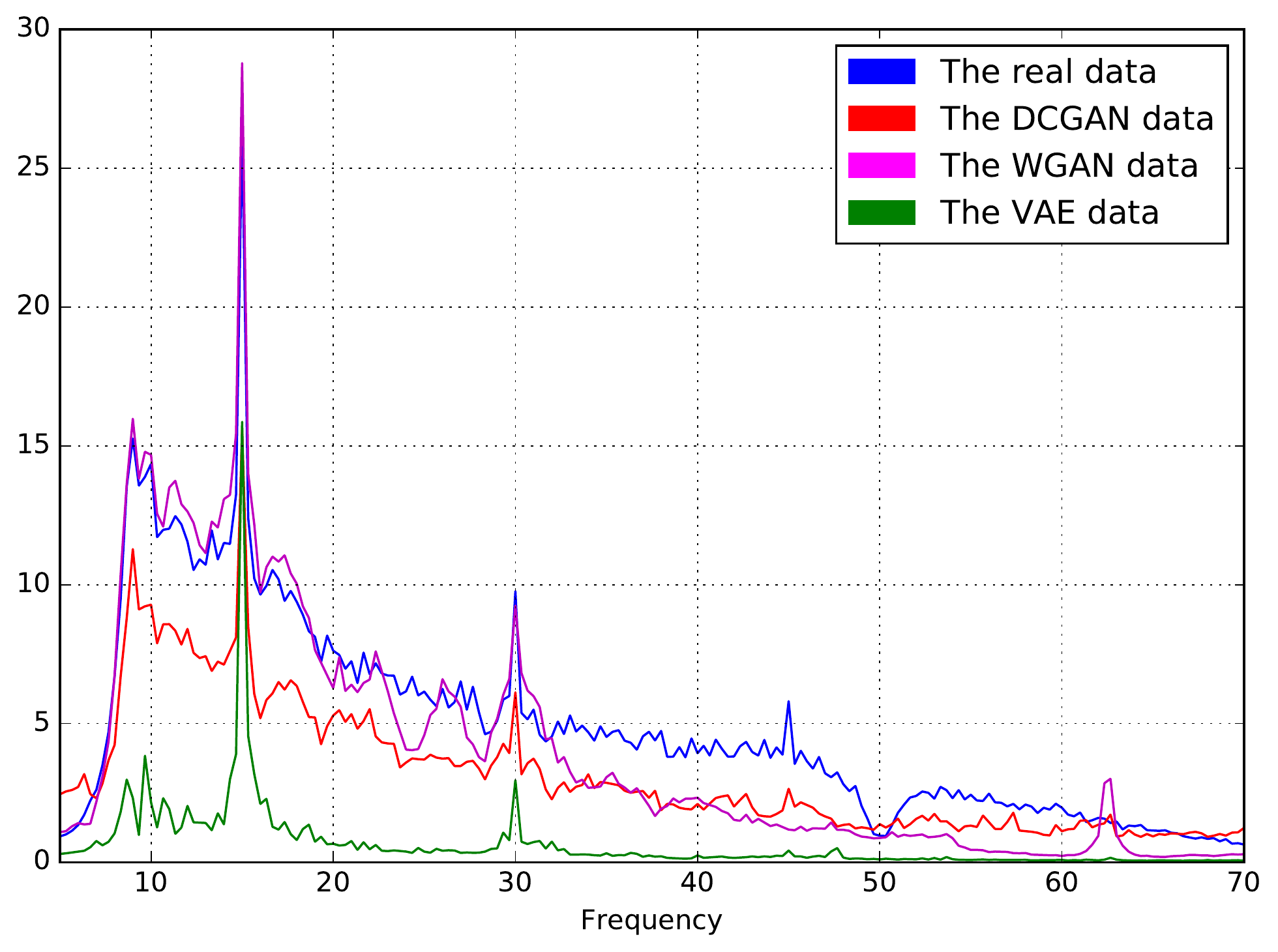}
      \vspace{-0.5cm} 
      \caption{SSVEP Signal at 15Hz.}
    \label{fig:class2}
  \end{subfigure}
  \vspace{0.05cm}
  \caption{Comparing real and synthetic data from the generative models. Synthetic data clearly displays the characteristic SSVEP frequency peaks at the same frequencies as those observed in the real data.}
  \label{fig:FFT}
  \vskip -15pt
\end{figure*}

\subsection{Synthetic Data Generation Procedure}
\label{sec:synth-data-gen}

Each of the generative models are trained upon data taken only from the \emph{Video-Stimuli dataset}, detailed in Section \ref{sec:empirical-datasets}. This will allow us to explore if data generated under one condition, can be used to improve the classification of data recorded under another. Once training is complete, the models are used to generate entirely new and synthetic datasets, with one dataset being made for each model. A potentially unlimited amount of data could theoretically be generated from our models. For our experiments, we generate 500 unique samples, each three seconds in length, for each of the three SSVEP frequencies present in the original \emph{Video-Stimuli dataset}. 

\subsection{Classification Procedure}
\label{sec:classification-prodedure}

In our experimental evaluation, we investigate exploiting the generated data to improve the classification accuracy on the \emph{NAO dataset}. To perform this classification, we employ a model based on the SSVEP Convolutional Unit (SCU) Convolutional Neural Network (CNN) architecture \cite{aznan2018classification} (Figure \ref{fig:CNN}). Prior work has shown this model to outperform traditional approaches and even time-series specific models like Recurrent Neural Networks (RNN) when classifying SSVEP EEG data \cite{aznan2018classification}. The classification model used for all experiments comprises of 1D convolutional layers, with batch normalization and max pooling. We first pre-process the EEG channels by referencing the data, applying a bandpass filter between 9 to 60 Hz, a notch filter at 50 Hz and finally normalizing the data between 0 and 1.

%% file: Sections/result.tex
\section{Results}

Our experimental evaluation is designed to firstly assess if it is indeed possible to generate realistic EEG data containing SSVEP signals and secondly, if this synthetic data can be used in a variety of ways to improve the classification accuracy on other real-world empirical datasets. All the classification results presented are the mean results from five different random seeds in order to test robustness and each experiment uses identical train/test splits for all runs to allow for direct comparisons to be made.

\subsection{Data Visualization via Fast Fourier Transforms}
\label{sec:data-viz}

SSVEP data has the phenomenon of frequency tagging, where the primary visual areas in human cortical oscillates to match the frequency of the fluctuating sinusoidal cycle of the SSVEP stimuli presented to the subject \cite{andersen2015driving}. EEG signals recorded during an SSVEP task will contain the target frequency clearly identifiable in the frequency domain \cite{liu2014recent}.

To validate that our generative models are indeed producing viable SSVEP signals, we visualise both the real and synthetic data via the Fast Fourier Transform (FFT) to decompose the EEG signals into the frequency domain. Figure \ref{fig:FFT} displays the frequency plot of the new synthetic data generated from our DCGAN, WGAN and VAE compared against the empirical data. As can be seen, all models accurately capture the characteristic SSVEP peaks at the target frequency and associated harmonics \cite{rao2013brain}, with the VAE producing signals with a comparatively lower amplitude. This result is very encouraging as it strongly suggests generative models can produce realistic SSVEP dry-EEG data. 

\subsection{Mixed Real and Synthetic Data Classification}
\label{sec:mix-classification}

To evaluate the applicability of synthetic EEG data being used to improve classification results, we combine the synthetic (using 30 samples per class) and real data into a single training set from which the classifier (see Section \ref{sec:classification-prodedure} for details) is trained. Table \ref{table:mixed data} displays the results, with values given per and across all subjects, where the baseline is no synthetic data included in the training set. It is also interesting to note that no single generative models data source performs the best across all subjects. 

\begin{table}[h!]
  \centering
  \resizebox{\columnwidth}{!}{
      \begin{tabular}{c c c c c c}
        \toprule
        \textbf{Method} & \textbf{S01} & \textbf{S02} & \textbf{S03} &\textbf{Across Subjects}\\
        \midrule \midrule
        \textbf{Baseline} & 0.91 $\pm$ 0.04   & 0.87 $\pm$ 0.10  & 0.84 $\pm$ 0.03 & 0.69 $\pm$ 0.03 \\
        \textbf{DCGAN} & 0.86 $\pm$ 0.01 & 0.80 $\pm$ 0.04 & \textbf{0.89} $\pm$ 0.03 & \textbf{0.72 $\pm$ 0.03} \\
        \textbf{WGAN} & \textbf{0.93 $\pm$ 0.04} & 0.90 $\pm$ 0.04  & 0.87 $\pm$ 0.02  & 0.71 $\pm$ 0.04   \\
        \textbf{VAE} &  0.92 $\pm$ 0.03 & \textbf{0.90 $\pm$ 0.03} & 0.79 $\pm$ 0.03 & 0.67 $\pm$ 0.02 \\
        \bottomrule 
      \end{tabular}}
    \caption{Classification test accuracy using generated and real-world data used to train the classifier. The baseline result is the classification of only real data.}
    \vskip -5pt
  \label{table:mixed data} 
\end{table}

Having a single classification model perform well across subjects is known to be highly challenging \cite{dehzangi2018portable}, yet possible \cite{aznan2018classification}. Table \ref{table:mixed data} illustrates how the inclusion of synthetic data within the training set can positively influence generalisation capabilities across subjects, which is an important result and of significant value in real-world applications. It is also interesting to note that although the synthetic data was generated from a different stimuli and sessions, its inclusion is still capable of improving the classification accuracy.

\subsection{Classification with Pre-Training}
\label{sec:pretrain-classification}

\begin{figure}[!h]
  \centering
  \begin{subfigure}[b]{0.28\textwidth}
    \includegraphics[width=\linewidth]{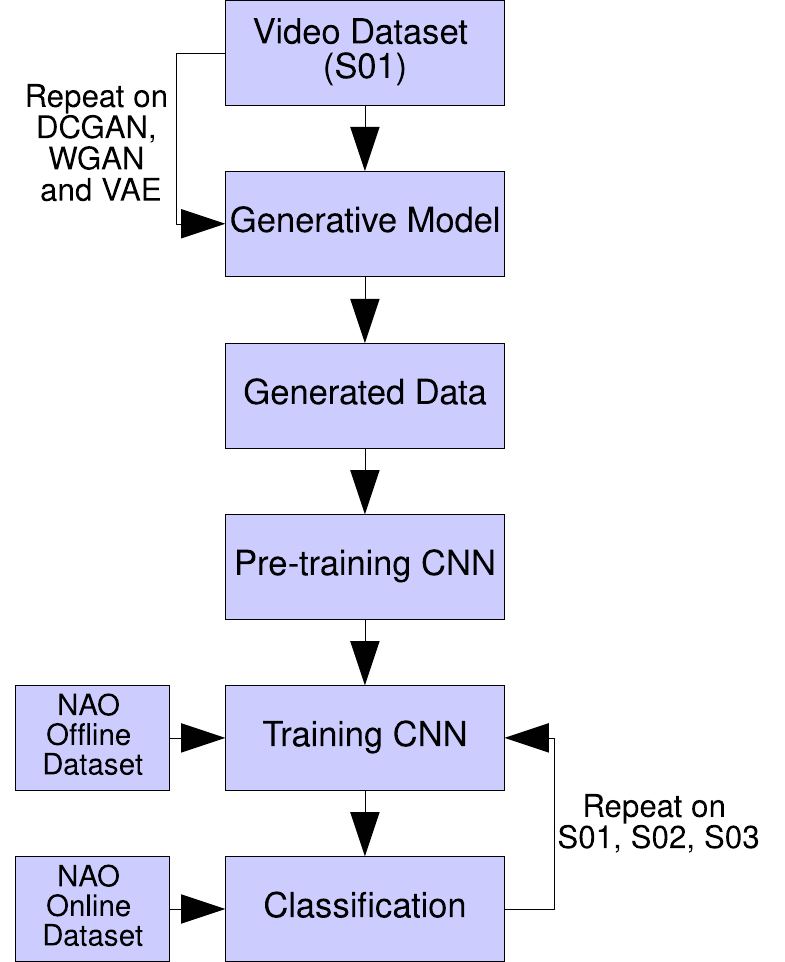}
      \caption{Pre-training classification \\ experiment.} 
      \label{fig:Flowchart_pretrain}
  \end{subfigure}
  \begin{subfigure}[b]{0.185\textwidth}
    \includegraphics[width=\linewidth]{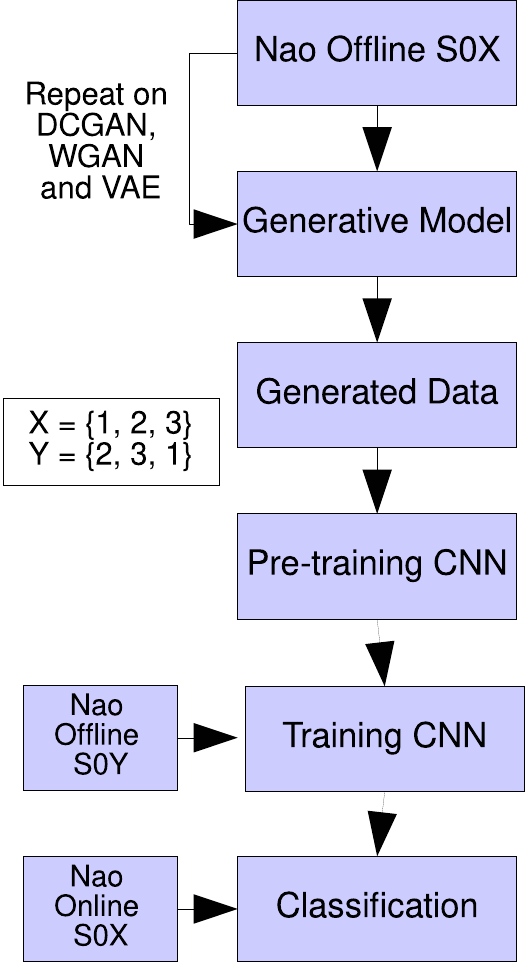}
    \caption{Training on data from different subjects.}
    \label{fig:Flowchart_different}
  \end{subfigure}
  \caption{Flowchart for experiment in section \ref{sec:pretrain-classification} (a) and experiment in section \ref{sec:cross-classification} (b).}
  \label{fig:experiment}
  \vskip -10pt
\end{figure}

\begin{figure}[h!]
  \centering
  \includegraphics[width=0.7\linewidth]{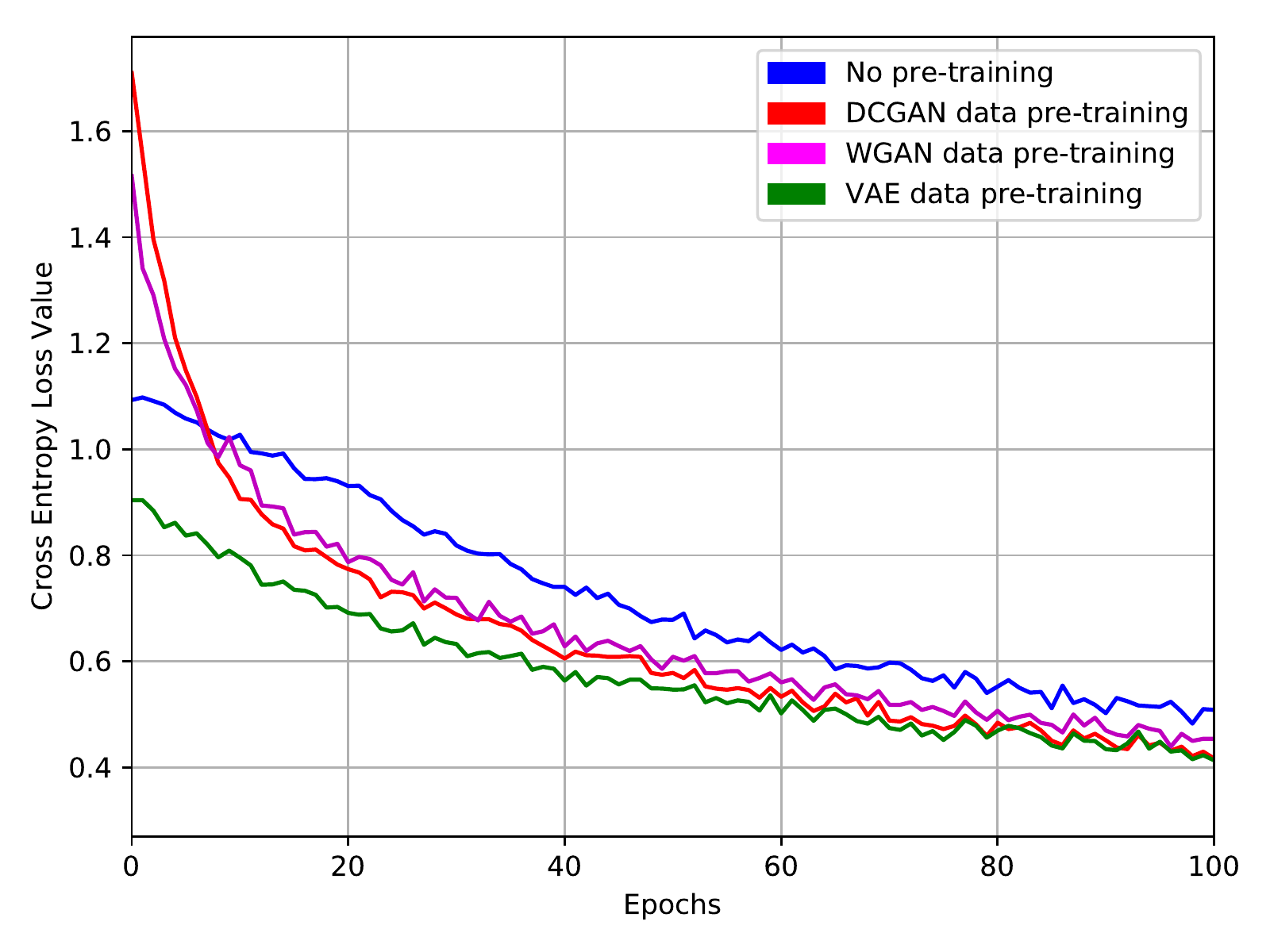} 
  \caption{Convergence of the Cross-Entropy value plotted over training epochs for models with and without pre-training on synthetic data.}
  \vskip -15pt
  \label{fig:loss_over_time}
\end{figure}

To further explore the usefulness of the synthetic data, we pre-train our classifier using only synthetic data (with 500 samples generated per class) and then further fine-tune the model using the real data as in Figure \ref{fig:Flowchart_pretrain}. The performance on the testing set is presented in Table \ref{table:pre-training}, where the baseline is no pre-training using synthetic data. Figure \ref{fig:loss_over_time} highlights how the loss values for this task vary over training epochs. It can be seen that models pre-trained using the synthetic data converge faster and to a lower overall loss value. 

\begin{table}[h!]
  \centering
  \resizebox{\columnwidth}{!}{ 
      \begin{tabular}{c c c c c c}
        \toprule
        \textbf{Method} & \textbf{S01} & \textbf{S02} & \textbf{S03} &\textbf{Across Subjects}\\
        \midrule \midrule
        \textbf{Baseline} & 0.91 $\pm$ 0.04   & 0.87 $\pm$ 0.10  & 0.84 $\pm$ 0.03 & 0.69 $\pm$ 0.03 \\
        \textbf{DCGAN} &   \textbf{0.97 $\pm$ 0.03} & \textbf{0.93 $\pm$ 0.03} & \textbf{0.87 $\pm$ 0.01} & 0.70 $\pm$ 0.02 \\
        \textbf{WGAN} &  0.93 $\pm$ 0.03 & 0.90 $\pm$ 0.04  & \textbf{0.87 $\pm$ 0.03}  & 0.72 $\pm$ 0.03 \\
        \textbf{VAE} &  0.93 $\pm$ 0.03  &  0.92 $\pm$ 0.06 & \textbf{0.87 $\pm$ 0.05} &  \textbf{0.73 $\pm$ 0.03}\\
        \bottomrule 
      \end{tabular}}
    \caption{Accuracy for test classification using synthetic data for pre-training stage. The baseline contains no pre-training.}
    \vskip -15pt
  \label{table:pre-training}
\end{table}

Both Table \ref{table:pre-training} and Figure \ref{fig:loss_over_time} demonstrate the ability to use the synthetic data to pre-train the network and improve the test accuracy on real data - another key result. This demonstrates the possibility of achieving higher classification accuracy using a smaller training set of real-world data, which resolves one of the most important challenges associated with SSVEP signal classification, unavailability of large datasets.

\begin{figure}[!h]
  \centering
  \begin{subfigure}[b]{0.23\textwidth}
    \includegraphics[width=\linewidth]{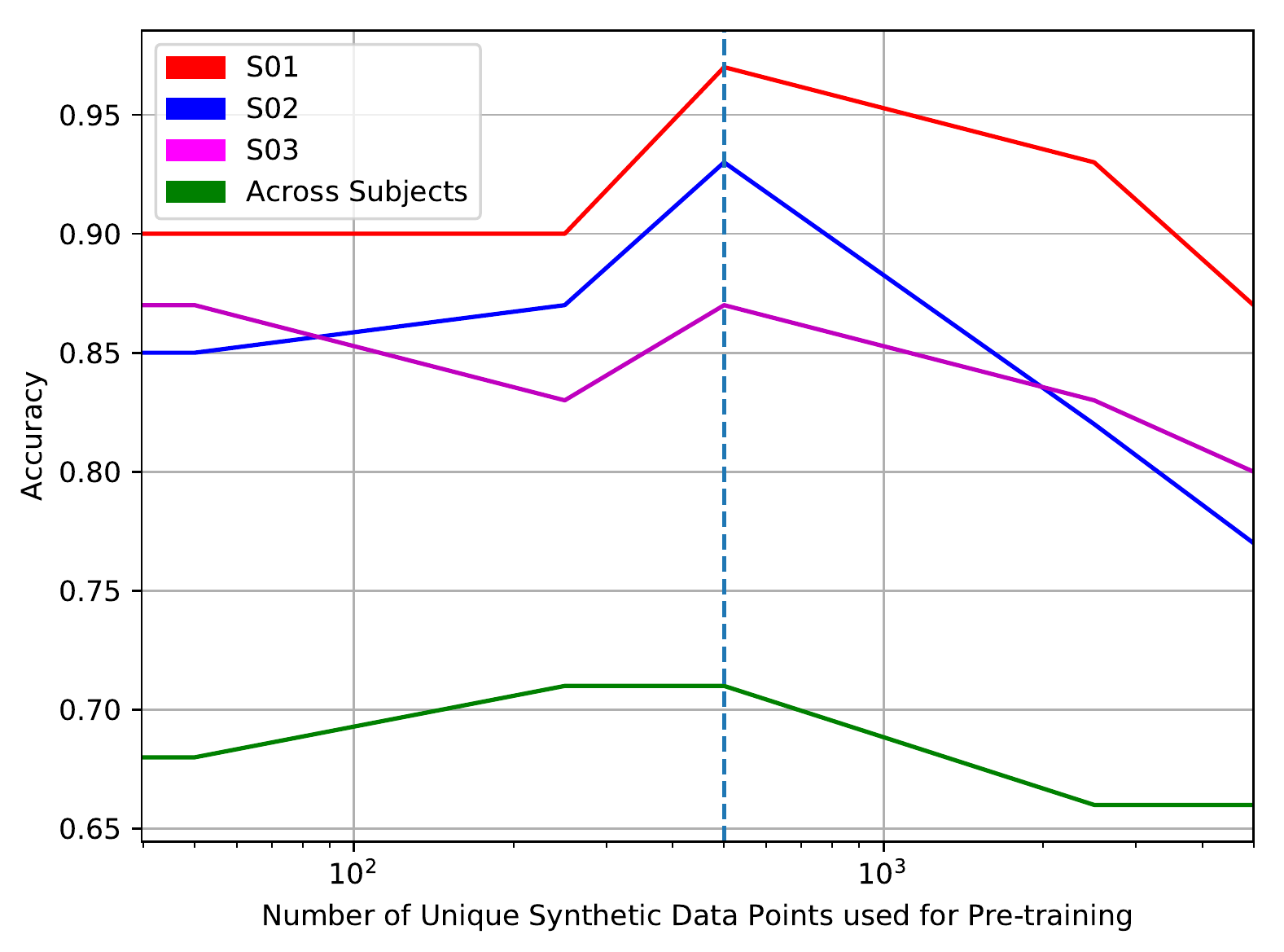}
      \caption{DCGAN} 
      \label{fig:DCGAN}
  \end{subfigure}
  \hfill
  \begin{subfigure}[b]{0.23\textwidth}
    \includegraphics[width=\linewidth]{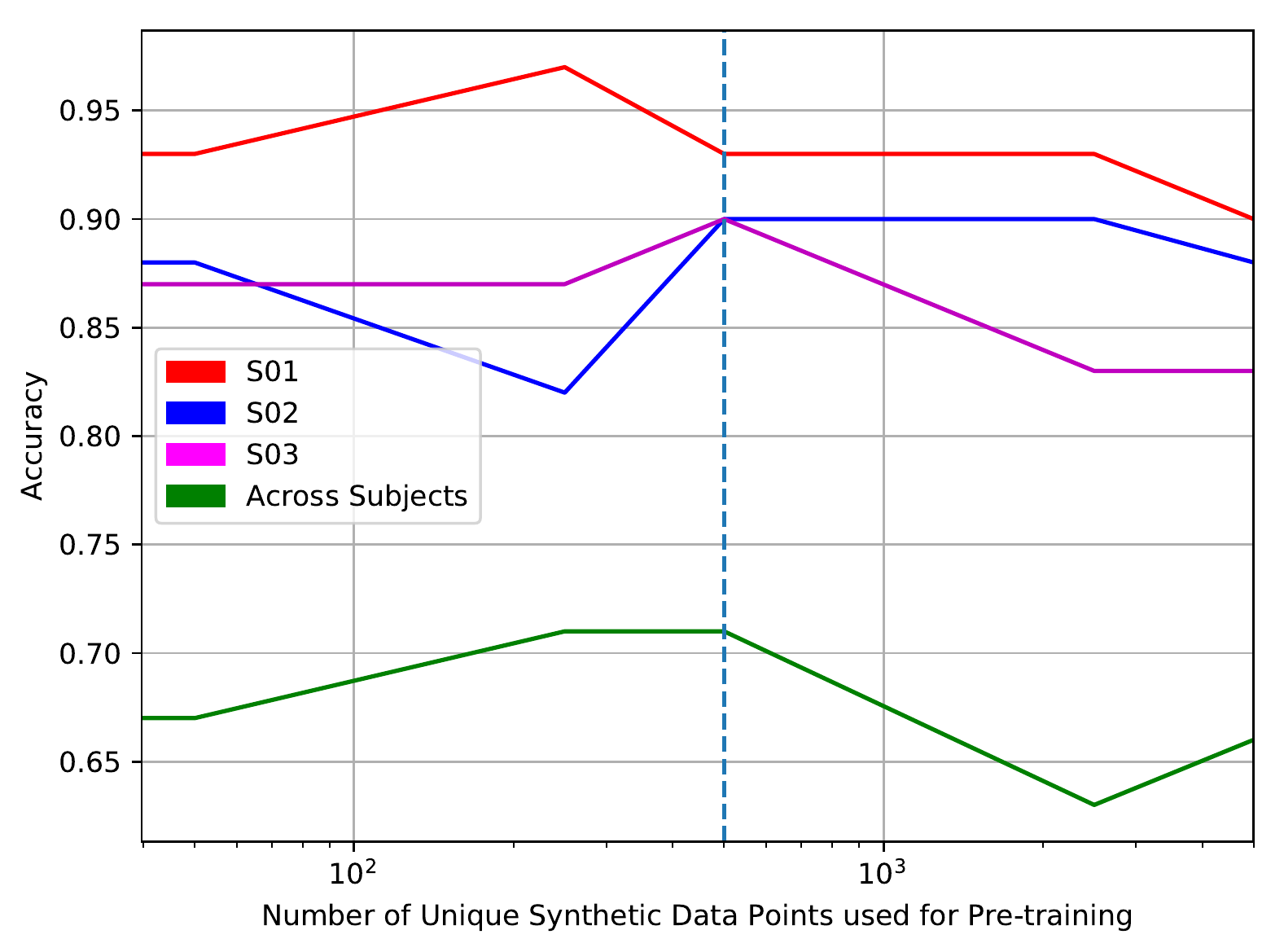}
      \caption{WGAN} 
      \label{fig:WGAN}
  \end{subfigure}

  \begin{subfigure}[b]{0.23\textwidth}
    \includegraphics[width=\linewidth]{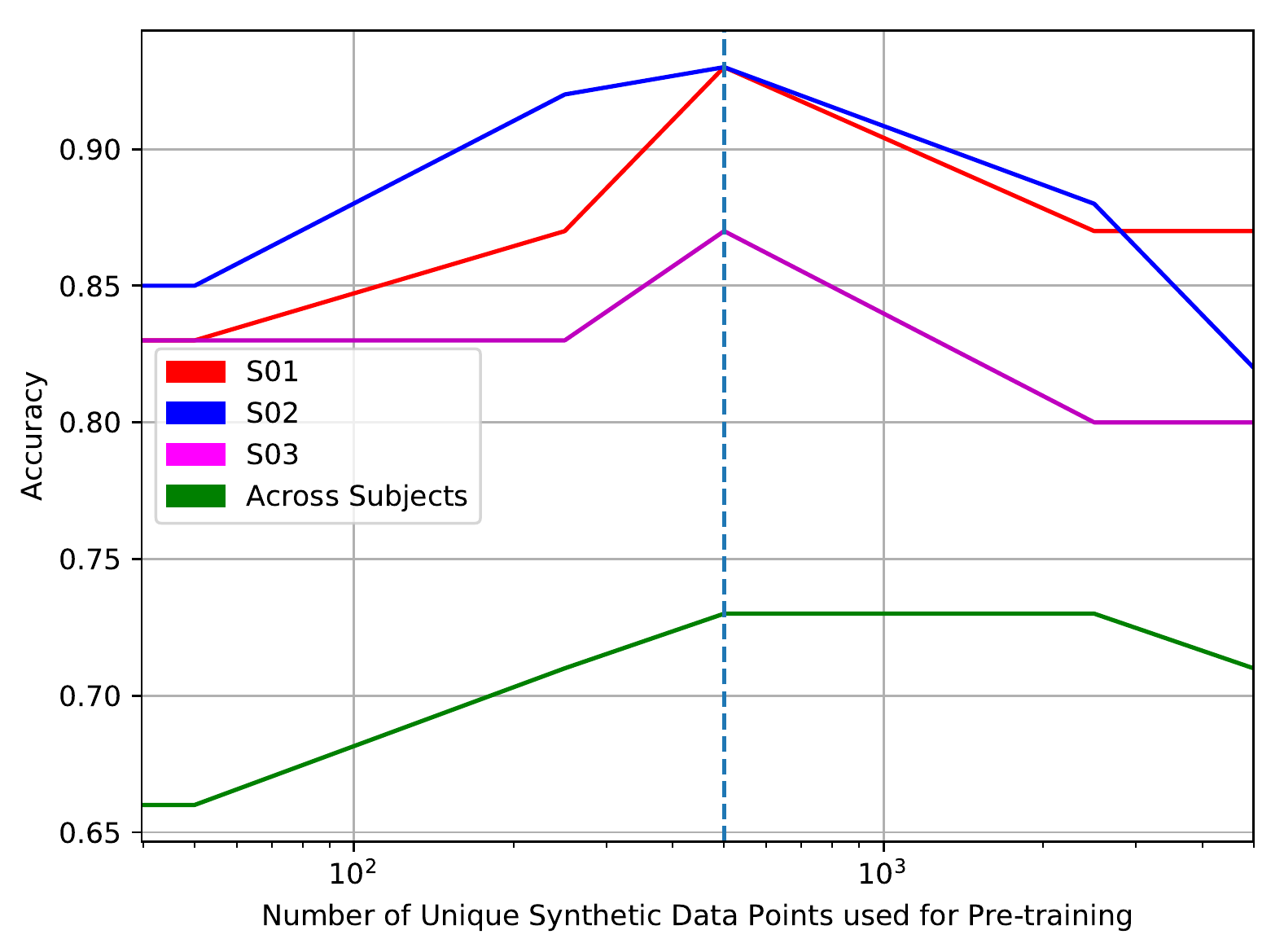}
      \caption{VAE}
      \label{fig:VAE}
  \end{subfigure}
  \begin{subfigure}[b]{0.23\textwidth}
    \includegraphics[width=\linewidth]{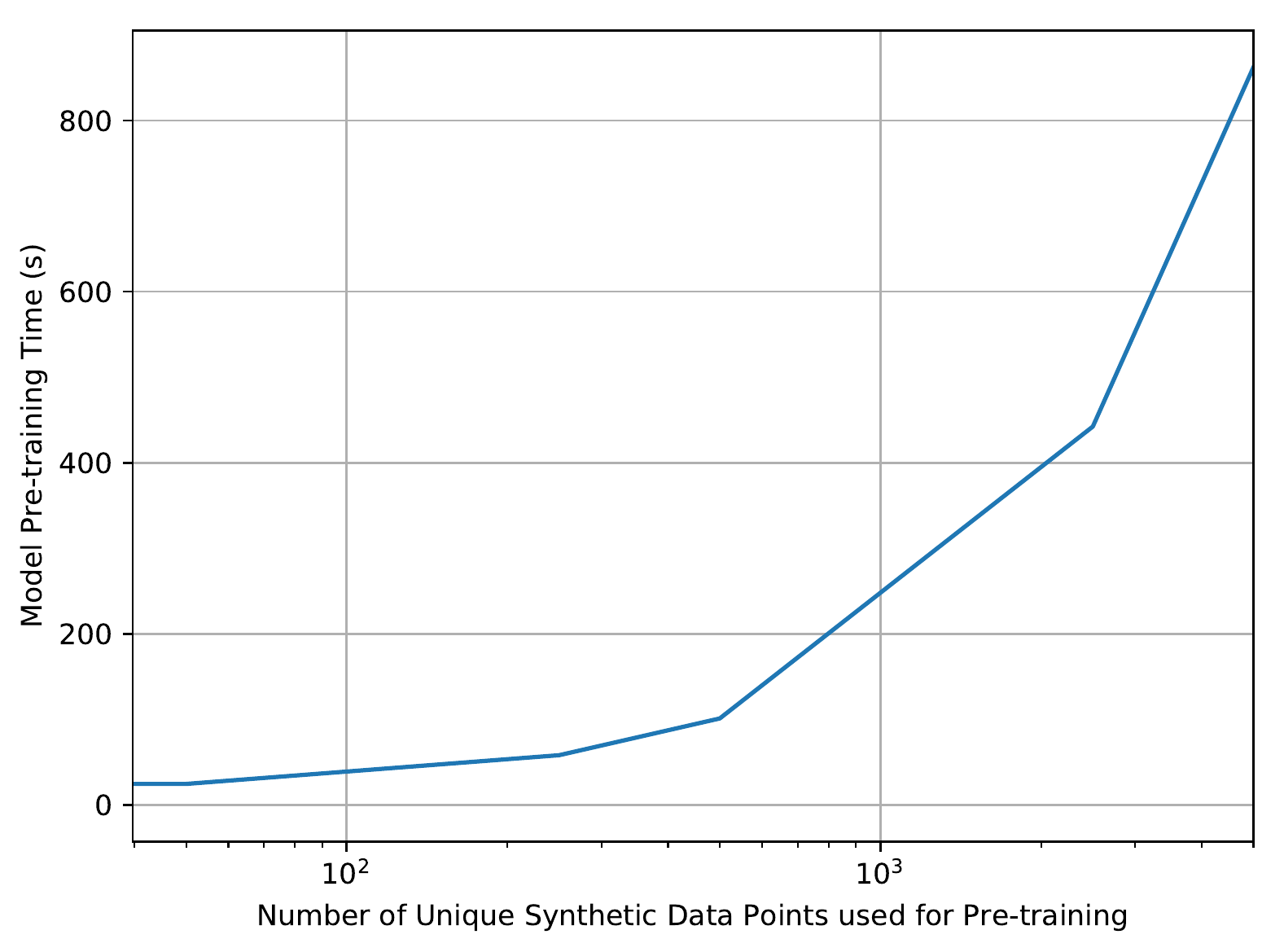}
      \caption{Pre-training time}
      \label{fig:Pretraintime}
  \end{subfigure}
  \caption{Test accuracy when varying the volume of synthetic data used for pre-training. Doted line indicates 500 samples.}
  \label{fig:data_quantity}
  \vskip -10pt
\end{figure}

Additional experiments were also conducted using varying quantities of synthetic data during pre-training. As seen in Figure \ref{fig:data_quantity}, the models perform better on average when 500 synthetic data samples are used in the pre-training stage. This is primarily due to the subject and session variant nature of dry-EEG data, since the models can easily over-fit to the distribution of the synthetic data in the presence of excessive pre-training. Furthermore, Figure \ref{fig:Pretraintime} demonstrates how the training time can rapidly increase when the size of the synthetic pre-training dataset grows too large. Accordingly, the use of 500 synthetic data points empirically offers an optimal trade-off between improved performance and tractable training time in this instance.

\subsection{Cross-Subject Generalisation}
\label{sec:cross-classification}

As mentioned in \ref{sec:intro}, SSVEP classification models often exhibit poor generalisation performance across different subjects or experiential data collecting sessions \cite{lotte2011generating, thomas2017deep}. In previous work \cite{aznan2018classification}, we successfully classified an unseen subject with no additional training required on that subject. In this section, we also attempt to demonstrate how pre-training on synthetic data can enhance the cross-subject generalisation capabilities of the model, i.e., the ability to classify data captured from one subject using a model trained on data captured from another. We train the generative models on S0X NAO Offline data (from the \emph{Nao Dataset}), pre-train the classifier using 500 samples of synthetic data output by said generative models, train the classifier on data from a different subject (S0Y) and finally test on online data from S0X from the \emph{Nao Dataset} (Figure \ref{fig:Flowchart_different}). As commonly seen within the literature \cite{lotte2018review, dehzangi2018portable}, one of the most important challenges in EEG-based research is the properties of EEG signals that vary from one subject to another as signal features can be specific to individual subjects.


\begin{table}[h!]
  \begin{center}
  \begin{tabular}{c c c c c c}
  \toprule
  \textbf{Method} & \textbf{S01} & \textbf{S02} & \textbf{S03} &\textbf{Mean}\\
  \midrule \midrule
  \textbf{Baseline} & 0.35 & 0.54  & 0.42 & 0.45 $\pm$ 0.08 \\
  \textbf{DCGAN} & 0.42 & 0.71  & 0.57 & 0.57 $\pm$ 0.12 \\
  \textbf{WGAN} & 0.40  & 0.60  & 0.50 & 0.56 $\pm$ 0.13 \\
  \textbf{VAE} & 0.70 & 0.90 & 0.85 & \textbf{0.82 $\pm$ 0.08} \\
  \bottomrule 
  \end{tabular}
  \end{center}
  \caption{Classification test accuracy for cross-subject Generalisation (see Figure \ref{fig:Flowchart_different}) via pre-training using synthetic data. The baseline does not include a pre-training stage.}
  \vskip -6pt
  \label{table:different subject}
  \end{table}

Table \ref{table:different subject} demonstrates how the baseline model performs poorly when trained using data collected from S0Y but tested on data from S0X, pointing to the lack of model Generalisability when EEG data is used. However, by pre-training the classifier on the generated data using NAO offline data from S0X, we see a large improvement in classification accuracy - especially when data generated by the VAE is used. This is a significant observation in the BCI research domain leading to a conjecture that, with further improvement in the results, we can eliminate the requirement for per-subject, per-session calibration for online applications.


%% file: Sections/conclusion.tex
\section{Conclusion}

In this work, we exploit recent advances made in neural-based generative models to explore potential benefits they can offer within the context of SSVEP classification models trained on EEG data. Since data acquisition within real-world scenarios suffers from a variety of challenges, using synthetically generated EEG signals can prove highly beneficial in improving the accuracy, convergence rate and generalization capabilities of any model trained to classify EEG data. We generate synthetic EEG signals using three state-of-the-art generative models - a Generative Adversarial Network, a Wasserstein Generative Adversarial Network and a Variational Auto-Encoder. Extensive evaluations demonstrate the efficacy of the synthetic data generated by said models across multiple experimental setups, with the inclusion of the generated data always improving the results. 

Future work will investigate the the influence of the quantity of the synthetic data generated using different approaches used during pre-training on the classification results can reveal valuable insight into the inner-workings of the classification process and nature of the synthetic data. Furthermore, we also plan to assess if mixing generated data taken from several models for pre-training can also improve our results.

%% file: ssvep-gan.bbl
\begin{thebibliography}{10}
\providecommand{\url}[1]{#1}
\csname url@samestyle\endcsname
\providecommand{\newblock}{\relax}
\providecommand{\bibinfo}[2]{#2}
\providecommand{\BIBentrySTDinterwordspacing}{\spaceskip=0pt\relax}
\providecommand{\BIBentryALTinterwordstretchfactor}{4}
\providecommand{\BIBentryALTinterwordspacing}{\spaceskip=\fontdimen2\font plus
\BIBentryALTinterwordstretchfactor\fontdimen3\font minus
  \fontdimen4\font\relax}
\providecommand{\BIBforeignlanguage}[2]{{%
\expandafter\ifx\csname l@#1\endcsname\relax
\typeout{** WARNING: IEEEtran.bst: No hyphenation pattern has been}%
\typeout{** loaded for the language `#1'. Using the pattern for}%
\typeout{** the default language instead.}%
\else
\language=\csname l@#1\endcsname
\fi
#2}}
\providecommand{\BIBdecl}{\relax}
\BIBdecl

\bibitem{rao2013brain}
R.~P. Rao, \emph{Brain-Computer Interfacing: an Introduction}.\hskip 1em plus
  0.5em minus 0.4em\relax Cambridge University Press, 2013.

\bibitem{lotte2018review}
F.~Lotte, L.~Bougrain, A.~Cichocki, M.~Clerc, M.~Congedo, A.~Rakotomamonjy, and
  F.~Yger, ``A review of classification algorithms for {EEG}-based
  brain--computer interfaces: a 10 year update,'' \emph{J. Neural Engineering},
  vol.~15, no.~3, p. 031005, 2018.

\bibitem{lotte2011generating}
F.~Lotte, ``Generating artificial {EEG} signals to reduce {BCI} calibration
  time,'' in \emph{Int. Brain-Computer Interface Workshop}, 2011, pp. 176--179.

\bibitem{kwak2017convolutional}
N.-S. Kwak, K.-R. M{\"u}ller, and S.-W. Lee, ``A convolutional neural network
  for steady state visual evoked potential classification under ambulatory
  environment,'' \emph{PloS one}, vol.~12, no.~2, p. e0172578, 2017.

\bibitem{aznan2018classification}
N.~K.~N. Aznan, S.~Bonner, J.~D. Connolly, N.~A. Moubayed, and T.~P. Breckon,
  ``On the classification of {SSVEP}-based dry-{EEG} signals via convolutional
  neural networks,'' in \emph{Int. Conf. Systems, Man, and Cybernetics}.\hskip
  1em plus 0.5em minus 0.4em\relax IEEE, 2018, pp. 3716--3721.

\bibitem{thomas2017deep}
J.~Thomas, T.~Maszczyk, N.~Sinha, T.~Kluge, and J.~Dauwels, ``Deep
  learning-based classification for brain-computer interfaces,'' in \emph{Int.
  Conf. Systems, Man, and Cybernetics}.\hskip 1em plus 0.5em minus 0.4em\relax
  IEEE, 2017, pp. 234--239.

\bibitem{zhang2018improving}
Q.~Zhang and Y.~Liu, ``Improving brain computer interface performance by data
  augmentation with conditional deep convolutional generative adversarial
  networks,'' \emph{arXiv preprint arXiv:1806.07108}, 2018.

\bibitem{luo2018eeg}
Y.~Luo, ``{EEG} data augmentation for emotion recognition using a conditional
  wasserstein {GAN},'' in \emph{Int. Conf. Engineering in Medicine and Biology
  Society}, 2018, pp. 2535--2538.

\bibitem{corley2018deep}
I.~A. Corley and Y.~Huang, ``Deep eeg super-resolution: Upsampling {EEG}
  spatial resolution with generative adversarial networks,'' in \emph{Int.
  Conf. Biomedical \& Health Informatics}.\hskip 1em plus 0.5em minus
  0.4em\relax IEEE, 2018, pp. 100--103.

\bibitem{salakhutdinov2015learning}
R.~Salakhutdinov, ``Learning deep generative models,'' \emph{Annual Review of
  Statistics and Its Application}, vol.~2, pp. 361--385, 2015.

\bibitem{goodfellow2014generative}
I.~Goodfellow, J.~Pouget-Abadie, M.~Mirza, B.~Xu, D.~Warde-Farley, S.~Ozair,
  A.~Courville, and Y.~Bengio, ``Generative adversarial nets,'' in
  \emph{Advances in Neural Information Processing Systems}, 2014, pp.
  2672--2680.

\bibitem{AdarDKAGG18}
M.~Frid{-}Adar, I.~Diamant, E.~Klang, M.~Amitai, J.~Goldberger, and
  H.~Greenspan, ``{GAN}-based synthetic medical image augmentation for
  increased {CNN} performance in liver lesion classification,''
  \emph{Neurocomputing}, vol. 321, pp. 321--331, 2018.

\bibitem{edlinger2012can}
G.~Edlinger and C.~Guger, ``{Can Dry EEG Sensors Improve the Usability of SMR,
  P300 and SSVEP Based BCIs?}'' in \emph{Towards Practical Brain-Computer
  Interfaces}, 2012, pp. 281--300.

\bibitem{Lopez-Gordo2014}
M.~A. Lopez-Gordo, D.~Sanchez-Morillo, and F.~{Pelayo Valle}, ``{Dry EEG
  electrodes},'' \emph{Sensors}, vol.~14, no.~7, pp. 12\,847--12\,870, 2014.

\bibitem{minguillon2017trends}
J.~Minguillon, M.~A. Lopez-Gordo, and F.~Pelayo, ``{Trends in {EEG-BCI} for
  Daily-life: Requirements for Artifact Removal},'' \emph{Biomedical Signal
  Processing and Control}, vol.~31, pp. 407--418, 2017.

\bibitem{liu2014recent}
Q.~Liu, K.~Chen, Q.~Ai, and S.~Q. Xie, ``Recent development of signal
  processing algorithms for {SSVEP}-based brain computer interfaces,'' \emph{J.
  Medical and Biological Engineering}, vol.~34, no.~4, pp. 299--309, 2014.

\bibitem{dehzangi2018portable}
O.~Dehzangi and M.~Farooq, ``Portable brain-computer interface for the
  intensive care unit patient communication using subject-dependent ssvep
  identification,'' \emph{BioMed research international}, vol. 2018, 2018.

\bibitem{aznan2018using}
N.~K.~N. Aznan, J.~D. Connolly, N.~A. Moubayed, and T.~P. Breckon, ``Using
  variable natural environment brain-computer interface stimuli for real-time
  humanoid robot navigation,'' in \emph{IEEE International Conference on
  Robotics and Automation}.\hskip 1em plus 0.5em minus 0.4em\relax IEEE, 2019.

\bibitem{jaakkola1999exploiting}
T.~Jaakkola and D.~Haussler, ``Exploiting generative models in discriminative
  vlassifiers,'' in \emph{Advances in Neural Information Processing Systems},
  1999, pp. 487--493.

\bibitem{arjovsky2017towards}
M.~Arjovsky and L.~Bottou, ``Towards principles for training generative
  adversarial networks,'' in \emph{Int. Conf. Learning Representations}, 2017,
  pp. 1--17.

\bibitem{arjovsky2017wasserstein}
M.~Arjovsky, S.~Chintala, and L.~Bottou, ``Wasserstein generative adversarial
  networks,'' in \emph{Int. Conf. Machine Learning}, 2017, pp. 214--223.

\bibitem{gulrajani2017improved}
I.~Gulrajani, F.~Ahmed, M.~Arjovsky, V.~Dumoulin, and A.~C. Courville,
  ``Improved training of wasserstein {GAN}s,'' in \emph{Advances in Neural
  Information Processing Systems}, 2017, pp. 5769--5779.

\bibitem{hinton2006reducing}
G.~E. Hinton and R.~R. Salakhutdinov, ``Reducing the dimensionality of data
  with neural networks,'' \emph{Science}, vol. 313, no. 5786, pp. 504--507,
  2006.

\bibitem{kingma2013auto}
D.~P. Kingma and M.~Welling, ``Auto-encoding variational bayes,'' \emph{arXiv
  preprint arXiv:1312.6114}, 2013.

\bibitem{rezende2014stochastic}
D.~J. Rezende, S.~Mohamed, and D.~Wierstra, ``Stochastic backpropagation and
  approximate inference in deep generative models,'' in \emph{Int. Conf.
  Machine Learning}, 2014, pp. 1278--1286.

\bibitem{larsen2015autoencoding}
A.~B.~L. Larsen, S.~K. S{\o}nderby, H.~Larochelle, and O.~Winther,
  ``Autoencoding beyond pixels using a learned similarity metric,'' in
  \emph{Int. Conf. Machine Learning}, 2016, pp. 1558--1566.

\bibitem{hartmann2018eeg}
K.~G. Hartmann, R.~T. Schirrmeister, and T.~Ball, ``{EEG-GAN}: Generative
  adversarial networks for electroencephalograhic ({EEG}) brain signals,''
  \emph{arXiv preprint arXiv:1806.01875}, 2018.

\bibitem{millan2004need}
J.~R. Millan, ``On the need for on-line learning in brain-computer
  interfaces,'' in \emph{Int. Joint Conf. Neural Networks}, 2004, pp.
  2877--2882.

\bibitem{palazzo2017generative}
S.~Palazzo, C.~Spampinato, I.~Kavasidis, D.~Giordano, and M.~Shah, ``Generative
  adversarial networks conditioned by brain signals,'' in \emph{Int. Conf.
  Computer Vision}, 2017, pp. 3410--3418.

\bibitem{schlogl2003outcome}
A.~Schl{\"o}gl, ``Outcome of the {BCI}-competition 2003 on the graz data set,''
  \emph{Berlin, Germany: Graz University of Technology}, 2003.

\bibitem{qiu2018data}
J.-L. Qiu and W.-Y. Zhao, ``Data encoding visualization based cognitive emotion
  recognition with {AC-GAN} applied for denoising,'' in \emph{Int. Conf.
  Cognitive Informatics \& Cognitive Computing}, 2018, pp. 222--227.

\bibitem{radford2015unsupervised}
A.~Radford, L.~Metz, and S.~Chintala, ``Unsupervised representation learning
  with deep convolutional generative adversarial networks,'' \emph{arXiv
  preprint arXiv:1511.06434}, 2015.

\bibitem{pytorch}
A.~Paszke, S.~Gross, S.~Chintala, G.~Chanan, Z.~DeVito, Z.~Lin, A.~Desmaison,
  and A.~Lerer, ``Automatic differentiation in {PyTorch},'' in \emph{Advances
  in Neural Information Processing Systems}, 2017, pp. 1--4.

\bibitem{kingma2014adam}
D.~Kingma and J.~Ba, ``Adam: A method for stochastic optimization,'' in
  \emph{Int. Conf. Learning Representations}, 2014, pp. 1--15.

\bibitem{liu2016ssd}
W.~Liu, D.~Anguelov, D.~Erhan, C.~Szegedy, S.~Reed, C.-Y. Fu, and A.~C. Berg,
  ``{SSD}: Single shot multibox detector,'' in \emph{Euro. Conf. Computer
  Vision}, 2016, pp. 21--37.

\bibitem{andersen2015driving}
S.~K. Andersen and M.~M. M{\"u}ller, ``Driving steady-state visual evoked
  potentials at arbitrary frequencies using temporal interpolation of stimulus
  presentation,'' \emph{BMC Neuroscience}, vol.~16, no.~1, p.~95, 2015.

\end{thebibliography}
